\documentstyle[12pt]{article}

\newcommand{\bcn}{\begin{center}}
\newcommand{\beq}{\begin{equation}}
\newcommand{\beqn}{\begin{eqnarray}}
\newcommand{\ecn}{\end{center}}
\newcommand{\eeq}{\end{equation}}
\newcommand{\eeqn}{\end{eqnarray}}

 \def\lsim{\mathrel{\rlap{\lower4pt\hbox{\hskip1pt$\sim$}}
    \raise1pt\hbox{$<$}}}
\def\slash#1{\setbox0=\hbox{$#1$}#1\hskip-\wd0\hbox to\wd0{\hss\sl/\/\hss}}
\begin{document}

\rightline{KU-HEP-93-28}
\bcn
\null
\null\vskip 1.5in
{\large EVIDENCE FOR OBSERVATION OF COLOR TRANSPARENCY IN pA COLLISIONS
USING GLOBAL FIT AND SCALING LAW ANALYSIS}\footnote{To appear in the
Proceedings of the XXVIII Rencontres de Moriond, QCD and High Energy
Interactions, Les Arcs (France, March 1993), J. Tranh Van, editor
(Editions Frontiers)}
\vspace*{1.0cm}

{\bf Pankaj Jain and John P. Ralston} \\

{\it Department of Physics and Astronomy \\
The University of Kansas \\
Lawrence, KS-66045-2151\\ }\vspace*{1.0cm}

\ecn

\vskip 2.0in
\noindent  ABSTRACT

\bigskip

\noindent
We review a new systematic data analysis procedure for color transparency
experiments and its application to the proton nucleus scattering experiment.
The method extracts the hard scattering rate as well as the
survival probability for the protons travelling through the nucleus
directly from the data. The method requires modelling of the nuclear
attenuation in terms of an effective nucleon-nucleon cross section.
We can minimize this model dependence by
introducing a new scaling law analysis procedure which can
yield considerable information with very little theoretical input.
With sufficient data the functional forms of the survival probability
as well as the hard scattering can be extracted without any theoretical
modelling.
An analysis of the BNL data of Carroll {\it et al} shows
clear evidence for observation of color transparency.
\vfill
\eject
\hspace*{3pc}
\baselineskip=18pt

\noindent
{\bf 1.} Color transparency$^{1)}$ is a theoretical prediction that
under certain
circumstances the strong  interactions may appear to be effectively reduced in
magnitude. Consider an exclusive
quasi-elastic process in which an incoming proton
knocks a proton out of a nucleus with large momentum transfer
without disturbing the rest of the nucleus.
If only short distance components of the
quark wave function in the proton contribute to this process,
then the incoming and outgoing states should have considerably reduced
attenuation as they propagate through the nucleus.
It would seem that to  make a
quantitative measurement of the attenuation, one must have a value for the hard
scattering  rate.  In the absence of a normalization of the hard sub-process,
 only a combination of  the hard scattering rate and
attenuation rate is measured in an experiment. This has lead to controversies
in the interpretation of the data.

\medskip

Theory at present cannot supply absolute normalizations for exclusive
processes, so the scattering  rate in an isolated hadron has been used
previously as a benchmark for comparison with the nuclear  target. The
``transparency ratio'' $T(Q^2,\ A)$
was introduced by Carroll et al$^{2)}$:  it is the
ratio of a  cross section measured in the nuclear target to the analogous cross
section for isolated hadrons in  free space. However it has become clear
that this ratio is not directly related to the attenuation
in the nuclear medium. Any soft components of the hadronic wave function
 contributing to the free space cross section should be
filtered by the nucleus, resulting in the survival of a short
distance part of the wave function$^{3)}$. Therefore the free space cross
section may be quite different from the hard scattering cross section
inside the nuclear target. If only the hard components contribute
to the nuclear hard scattering, we expect that it will be closer
to the perturbative QCD (pQCD) predictions. In contrast, the free space
fixed angle $pp\rightarrow pp$ cross
section is known to deviate considerably from pQCD, displaying, e.g.
oscillations with energy.

\medskip

In order to avoid making too many assumptions about the hard scattering
we recently introduced$^{4)}$ a more systematic data analysis procedure which
does not presume that the hard scattering rate inside the nucleus
is known. The procedure
is based on fitting the A dependence of the nuclear cross section
$d\sigma_A$ at fixed momentum transfer ($Q^2$) to determine the effective
attenuation
cross section $\sigma_{eff}$. Assuming that the hard scattering
factorizes from the subsequent propagation of the hadron, a transparency
ratio $T(Q^2,A)$ can be written as

\beqn
T(Q^2,A)&=&{d\sigma/ dt|_{hard,\ A}\over Z\ d\sigma/ dt|_{free\ space}}\
 P(\sigma_{eff}(Q^2),\; A)\nonumber\\
&=& f(Q^2)\ P(\sigma_{eff}(Q^2),\; A)
\eeqn

\noindent
where $P$ is the survival probability. Here $f(Q^2)$ is
treated as an unknown function
which contains information about the hard scattering inside the
nuclear target as well as the free space scattering. For the BNL fixed target
experiment we take
 $Q^2= -t\approx s/2$.
By fitting the $A$ dependence of $d\sigma_A$ at fixed $Q^2$ we
determine the functional dependence of $P$ on $A$.
In our application to the
BNL $pA\rightarrow p'p''(A-1)$ experiment we used a semiclassical
model for the survival probability
 which assumes exponential attenuation of the protons with
effective cross section $\sigma_{eff}$ as they propagate through
the nucleus.

By following this procedure we can determine $\sigma_{eff}(Q^2)$ and
$f(Q^2)$ at each $Q^2$. Since $\sigma_{eff}(Q^2)$ and
the normalization $f(Q^2$) are free, the fit to  the data may or may not show
that these parameters vary with energy. The global fit$^{4}$ to the BNL
data shows evidence for observation of color transparency: $\sigma_{eff}
=17\pm 2\ mb\ (12\pm 2\ mb)$ at the $Q^2$value of 4.8 $GeV^2(8.5\ GeV^2)$.
By use of the intermediate $Q^2$ data points reported for the
Aluminum target we determine $\sigma_{eff}\cong 2.2\ GeV^2/Q^2$, a
rate of decrease in agreement with the pQCD predictions$^{5)}$. The plot of
$\sigma_{eff}$
deduced from data is shown in Fig. (1).
 The fit also shows that the hard scattering
inside the nuclear medium goes down with energy at
a rate faster than $(Q^2)^{-10}$. Instead, the hard scattering rate is in
agreement with the pQCD prediction
which goes roughly like $\alpha_s^{10}(Q^2)/(Q^2)^{10}$.

The above procedure can be made even more model independent by
using the scaling law$^{6)}$ for data analysis. As
introduced in Ref. [6],  the scaling law was assumed to
apply in electroproduction to the transparency ratio. (At that time it was
believed that one had to assume the hard scattering rate cancelled out in the
ratio.) Now it is clear that
 the scaling law is instead
 valid for the survival probability. Basically it says that the
survival probability is a function of a dimensionless variable.
The important dimensionless variable that it can depend on is the
effective
number of nucleons encountered by the protons as they propagate
through the nucleus.
This is proportional to the length of the nucleus $(A^{1/3})$
times the nuclear density $n$ times the effective nucleon-nucleon
 cross section $\sigma_{eff}$. If the cross section goes
like $1/Q^2$ then the
survival probability is a function of the dimensionless
quantity $n A^{1/3}/Q^2$.

We need to
experimentally determine whether this law is satisfied or not.
We consider the survival probability to be a function of
$Q^2/A^\alpha$ and determine $\alpha$ from the experimental data.
The basic complication in this determination is the
fact that the experimentally measured transparency ratio $T$
involves an unknown function $f(Q^2)$,
where $f(Q^2)$ is the ratio of the hard scattering in the nuclear
medium to the free space scattering. With a good data set over a range of
the $Q^2$ and $A$ plane one can nevertheless extract functional
forms for both $f(Q^2)$ and the survival probability $P$ by fitting
$T=f(Q^2)P(Q^2/A^\alpha)$. This is attractive because of its
model independence: no theory needs to be used to model $P$, which is
simply determined by the best fit.
However,
 the present data is limited since it is available only at
two energies for several $A$. We therefore take a somewhat
different approach here to check the scaling law with the present
data. We assume that the hard scattering inside the nuclear
medium satisfies the short distance pQCD predictions. This is
based on the idea of nuclear filtering and is supported by
the global fit$^{4)}$ and by the fact that the cross section in the
nuclear target does not show any oscillations$^{3}$. We therefore set the
average
hard
scattering cross section in eq. (1) to be Z times the short distance
perturbative QCD prediction$^{7)}$, which is given by

\beq
d\sigma/dt_{pQCD} \cong
(\alpha_s(Q^2/\Lambda_{qcd}^2))^{10}s^{-10}f(t/s)\ .
\eeq

\noindent
(In the above expression we have ignored the
anomalous dimension which is found to make a very small difference
in the results.)
We can then calculate
$f(Q^2)$ and take it out of the transparency ratio to yield
the survival probability.
The survival probabilities for three
different values of $\alpha$ are plotted in fig. 2. The data
points are circles $(Q^2=4.8\ GeV^2)$, squares $(Q^2=8.5\ GeV^2)$
and crosses $(Q^2= 10.4\ GeV^2)$. It
is clear from the figure that $\alpha\ =\ 1/3$ is favored over other
values of $\alpha$. This clearly shows that the data satisfies the
scaling law which can therefore be used effectively for analysing
future color transparency experiments.

There remains one subtlety. Although the scaling law procedure is
quite explicit about the functional dependence on the two variables
$Q^2$ and $Q^2/A^\alpha$, the ansatz (1) has a symmetry: the data and
fit are unchanged under $f(Q^2)\rightarrow f(Q^2)/K$, $P(Q^2/A^\alpha)
\rightarrow KP(Q^2/A^\alpha)$, where $K$ is any constant.
The absolute
magnitudes of the survival probability (or the hard scattering)
and the effective attenuation cross section are therefore
not determined by
this method.

In summary, a systematic global fit shows evidence for observation
of color transparency. Our results also show that the data obeys
the expected scaling behavior. It will clearly be of great interest
to analyze future color transparency experiments by using both methods.

\bigskip

\noindent
{\bf Acknowledgements:}    This work has been supported in part by the DOE
Grant No.
DE-FG02-85-ER-40214.A008.

\bigskip

\noindent {\bf References}
\bigskip
\begin{enumerate}
\item S. J. Brodsky and A. H. Mueller, Phys. Lett. B {\bf 206}, 685 (1988), and
references therein.

\item  A. S. Carroll et al., Phys. Rev. Lett. {\bf 61}, 1698 (1988).

\item J. P. Ralston and B. Pire, Phys. Rev. Lett. {\bf 61}, 1823 (1988);
ibid {\bf 65}, 2343 (1990).

\item P. Jain and J. P. Ralston, submitted to Phys. Rev. D; to appear
in the Proceedings of the Workshop on the {\em Future Directions in Nuclear and
Particle Physics at Multi GeV Energies}, edited by D. Geesaman et al.
(Brookhaven National Laboratory 1993) (in press).

\item S. Nussinov, Phys. Rev. Lett {\bf 34}, 1286 (1975); F. E. Low, Phys. Rev.
{\bf D12}, 163 (1975);  J.  Gunion and D. Soper, Phys. Rev. {\bf D15}, 2617
(1977).

\item B. Pire  and J. P. Ralston, Phys. Lett. {\bf 256}, 523 (1991).

\item S. J. Brodsky and G. P. Lepage, Phys. Rev. {\bf D22}, 2157 (1980).
\end{enumerate}

\end{document}